\documentclass[12pt]{article}
\begin{document}
\pagestyle{empty}

\begin{center}
{\Large \bf

Deep inelastic sum rules  at the boundaries between perturbative 
and non-perturbative QCD}

\vspace{0.1cm}

{\bf A.L. Kataev}\footnote{E-mail:kataev@ms2.inr.ac.ru }\\
\vspace{0.1cm}

Institute for Nuclear Research of the Russian  Academy of Sciences
,\\ 117312  Moscow, Russsia

\end{center}
\begin{center}

{\bf Abstract}
\end{center}
The basis of renormalon calculus is briefly discussed. This method is 
applied to study the  QCD predictions for three different sum rules 
of deep-inelastic scattering, namely for the Gross--Llewellyn Smith,
Bjorken polarized and unpolarized sum rules.
It is shown that the renormalon structures  of these a posteriori 
different physical quantities are closely related. 
These  properties are giving us 
the hint that theoretical expressions of these three sum rules 
are similar both in the perturbative and non-perturbative 
sectors. Some phenomenological consequences of 
the new  relations are discussed. 

\noindent 
\vspace*{0.1cm}
\noindent
\vspace*{2cm}

\vfill\eject

\setcounter{page}{1}
\pagestyle{plain}

\section{Introduction}
The formulation of the 
quantitative method  of renormalon calculus on the higher 
level of understanding  
takes its start from 
the important works of Ref.\cite{Zakharov:1992bx, Mueller}, devoted to the 
consideration of $e^+e^-\rightarrow{hadrons}$ process, and from  
the interesting  work of Ref.\cite{Mueller:1993pa}, devoted to 
the consideration of deep-inelastic scattering processes. 

After these studies  the number of  theoretical and practical 
developments appeared in the literature (see reviews of Refs. 
\cite{Beneke:1998ui}--\cite{Altarelli:1995kz}). 

In what is discussed below,  we  
will consider  aspects of renormalon calculus, 
related to   deep-inelastic scattering (DIS) sum rules.    

It is commonly expected that in the canonical renormalization 
schemes, say the $\overline{\rm MS}$ scheme, perturbative expansions in 
small QCD  coupling constant $a_s=\alpha_s/(4\pi)$ of theoretical expressions 
for physical quantities, defined in the Euclidean region, 
are asymptotic ones. This means that the difference of the total 
sums 
\begin{equation}
D(a)=1+\sum_{n\geq 1}d_na^n
\end{equation}
and their finite sums
\begin{equation}
D_{\rm k}(a)=1+\sum_{n=1}^{\rm{k}}d_na^{\rm k}
\end{equation}
satisfy the following property 
\begin{equation}
{\rm lim}_{a\rightarrow 0}|\frac{D(a)-D_{\rm{k}}(a)}{a^{\rm k}}
|\rightarrow 0~~~.
\end{equation}
In other words the difference between the total series and their
finite sum are expressed as
\begin{equation}
D(a)-D_{\rm k}(a)=O(a^{\rm{k}+1})~~~.
\end{equation}
In this case the  error of the truncation of the asymptotic series 
can be estimated by the last term  of  $D_{\rm{k}}(a)$ , namely 
$d_{\rm{k}}a^{\rm{k}}$ \cite{Dingle}.

In QCD one expects that in the $\overline{\rm MS}$ scheme 
the  coefficient function 
for DIS  sum rules, normalized to unity can be approximated  
by the following   asymptotic series\cite{Beneke:1998eq}:
\begin{equation}
D_{\rm{k}}(a)=1+\sum_{\rm {k}\geq 1}(\beta_0)^{\rm k} {\rm k}!
\bigg(K_D^{\rm {UV}} (-1)^{\rm k}{\rm {k}}^a+K_D^{\rm{IR}}{\rm k}^b\bigg)
a_s^{\rm{k+1}}~,
\label{asymptotics}
\end{equation}
where sign-alternating series with the coefficient 
 $K^{\rm{UV}}$ is generated by the ultraviolet renormalons (UVR), 
sign-constant asymptotic series with coefficient 
$K^{\rm{IR}}$ result from the  consideration of infrared renormalons (IRR),  
and $a$ and $b$ are the known numbers, that  depend from the 
ratio of the first two coefficients of the QCD $\beta$-function.

Working within  renormalon 
calculus we will demonstrate that 
the perturbative and non-perturbative contributions to  
definite DIS sum rules are related. 
In other words we will show  that the renormalon approach is working  
 at the boundaries between these two 
regimes in QCD.     

The aim of this article  is three fold:
\begin{itemize} 
\item to explain the basic stages of renormalon calculus in QCD,  
using the  simple language;
\item  
to show that in the asymptotic perturbative expansion 
of three DIS  sum rules, 
namely of the   
Gross--Llewellyn Smith (GLS), Bjorken- polarized (Bjp) and 
Bjorken- unpolarized (Bjunp) sum rules,   
may be universal.  
We will present arguments, based on the consideration of the results 
given  in Refs.\cite{Broadhurst:1993ru},\cite{Broadhurst:2002bi},
that these expansions are  defined by the poles in the closely 
related  Borel images 
of all three sum rules. 
\item 
We will explain the features,  which follow from the consideration  
of the IRR  poles in the  
Borel images of the   three DIS sum rules. Moreover our aim is 
to outline  new  
consequences of the IRR calculus. They  indicate  the 
existence of  relations between 
twist-4 
$1/Q^2$ non-perturbative contributions to the  sum rules 
we are interested in\cite{Kataev:2005ci}. 
 These results form the basis of   
the  new QCD relations between theoretical expressions 
for these three sum rules\cite{Kataev:2005ci}, which 
seem to be  supported by the experimental 
data within existing  error bars. 
More critical tests of these relations are proposed.

\end{itemize}
   
\section{Renormalon calculus and DIS sum rules}
%
%

Let us first  express 
a perturbative QCD  series in terms of a Borel integral
as 
\begin{eqnarray}
D(a_s)&=&\sum_{n=0}^{\infty}d_n a_s^{n} \\ \nonumber 
&=& \sum_{n=0}^{\infty} d_n\frac{n !}{\Gamma(n+1)}\delta^{n} \\ \nonumber 
&=& \int_0^{\infty} {\rm exp}(-\delta/\beta_0 a_s)
\sum_{n=0}^{\infty}d_n\frac{\delta^{n}}{n !}d\delta
\\ \label{Borel}
&=& \int_0^{\infty} {\rm exp} (-\delta/\beta_0 a_s) B[D]({\delta})d\delta~,`
\end{eqnarray}
where  $\beta_0=(11/3)C_A-(4/3)T_fN_f$ is the first 
coefficient of the QCD $\beta$-function, with $C_A=3$, $T_f=1/2$, and 
$B[D](\delta)$ is the image of the Borel integral. 

At this stage we define the DIS  sum rules 
we will be  interested in. The GLS sum rule of the  $\nu N$ 
DIS\cite{Gross:1969jf} has the following form   
\begin{eqnarray}
\nonumber
{\rm GLS}(Q^2)&=&\frac{1}{2}\int_0^1dx\bigg[F_3^{\nu n}(x,Q^2)+
F_3^{\nu p}(x,Q^2)\bigg] \\ 
&=&3 C_{\rm GLS}(Q^2)-\frac{\langle\langle O_1 \rangle\rangle}{Q^2}-
O\bigg(\frac{1}{Q^4}\bigg)~~. 
\end{eqnarray}
In the Born approximation, this  
``measures'' the number of valence quarks,  that  are contained 
in the nucleon and can thus  be considered as the  {\bf baryon sum rule}.
In the $\overline{\rm MS}$ scheme,  the twist-2 
 perturbative coefficient function 
$ C_{GLS}(Q^2)$ is calculated explicitly, including $a_s^2$
and $a_s^3$ terms\cite{Gorishnii:1985xm},\cite{Larin:1991tj}.
The twist-4 matrix element of the $O(1/Q^2)$ non-perturbative 
contribution  
to the GLS sum rule is related to the matrix element calculated 
in Ref.\cite{Shuryak:1981kj} to be 
\begin{equation}
\langle\langle O_1 \rangle\rangle=
\frac{8}{27}\langle\langle O^{\rm S}\rangle\rangle~~,  
\end{equation}
where  $\langle\langle  O^{\rm S}\rangle\rangle$ is defined by   
the  following operator 
\begin{equation}
O_{\mu}=
\overline{u}\tilde{G}_{\mu\nu}\gamma_{\nu}\gamma_5u+ (u\rightarrow d)~~, 
\end{equation}
where 
\begin{equation}
\label{def1}
\tilde{G}_{\mu\nu}=\frac{1}{2}\epsilon_{\mu\nu\alpha\beta}G_{\alpha\beta}^a
\frac{\lambda^a}{2}
\end{equation}
and 
\begin{equation}
\langle P|O_{\mu}^{\rm S}|P\rangle =2p_{\mu}\langle\langle O^{\rm S}
\rangle\rangle~~~. 
\end{equation}

The second sum rules, actively studied both in theory and experiment,
is the Bjp sum rule\cite{Bjorken:1966jh}, having  the physical meaning of 
{\bf polarized isospin sum rule}. Its   theoretical expression 
can be defined as 
\begin{eqnarray}
\nonumber
{\rm Bjp}(Q^2)&=&\int_0^1dx\bigg[g_1^{lp}(x,Q^2)-g_1^{ln}(x,Q^2)\bigg] \\ 
&=&
\frac{g_A}{6} C_{\rm Bjp}(Q^2)-\frac{\langle\langle O_{2} 
\rangle\rangle}
{Q^2}-O\bigg(\frac{1}{Q^4}\bigg)~~.
\end{eqnarray}
Here  $g_A=1.26$ is the known $\beta$-decay constant. 
At the $a_s^3$ level its perurbative part differs from the 
one of the GLS sum rule by the absence of small ``light-by-light''-type 
 terms, 
proportional to the colour structure $d^{abc}d^{abc}$ 
\cite{Larin:1991tj}. The  structure of the power corrections 
 to the matrix element of the 
 leading $O(1/Q^2)$ 
 power correction was 
analytically calculated in Ref.\cite{Shuryak:1981pi}, with the 
useful  correction  input from the considerations 
of Ref.\cite{Ji:1993sv}.  The final  expressions are  presented 
in a simple-form in the  review of Ref.\cite{Hinchliffe:1996hc}, 
from which we can get:  
\begin{equation}
\langle \langle O_2\rangle\rangle =\frac{1}{6}\frac{8}{9} \bigg[
\langle\langle  U^{\rm NS}\rangle\rangle -\frac{M_N^2}{4}
\langle\langle V^{\rm NS}\rangle
\rangle\bigg] ~~,
\end{equation}
where     
\begin{eqnarray}
\nonumber 
\langle P,S|U_{\mu}^{\rm NS}|P,S\rangle &=& 2M_N S_{\mu}\langle\langle 
U^{\rm NS}
\rangle\rangle 
\\ 
\langle P,S|V_{\mu\nu,\sigma}^{\rm NS}|P,S\rangle &=& 2M_N\langle\langle 
V^{\rm NS} \rangle\rangle\{(S_{\mu}P_{\nu}-S_{\nu}P_{\mu})P_{\delta})
\}_{S\{\nu,\sigma\}}
\end{eqnarray}
and $\langle\langle U^{\rm NS}
\rangle\rangle$ and  $\langle\langle V^{\rm NS}
\rangle\rangle$ 
are the  reduced matrix elements of the local operators from 
Ref.\cite{Shuryak:1981pi}, namely 
\begin{eqnarray}
\nonumber
U_{\mu}^{\rm NS}&=&g_s\big[\overline{u}\tilde{G}_{\mu,\nu}\gamma^{\nu}u 
-(u\rightarrow d)\big] \\
V^{\rm NS}_{\mu\nu,\sigma}&=&g_s\{\overline{u}\tilde{G}_{\mu\nu}\gamma_{\delta}
u - (u\rightarrow d)\}_{S\{\nu,\delta\}}~~~,
\end{eqnarray}
where  $S\{\nu,\sigma\}$ stand for symmetrization over the given subscripts
and $\tilde{G}_{\mu,\nu}$ is defined in Eq. (\ref{def1}).
In Ref.\cite{Balitsky:1989jb} the definition  of Eq. (14) was used for 
the estimates of $O(1/Q^2)$ 
corrections to Bjp sum rule, using the three-point function QCD  sum rules 
technique. These calculations were then re-analyzed 
with the same method in Ref.\cite{Ross:1993gb}. 
The numerical results of these calculations will be discussed  later.
In the work of Ref.\cite{Stein:1994zk} a similar 
analysis was done with the help  of the same   method for
the first term in the r.h.s. of Eq. (14), while the term, proportional 
to $(M_N^2/4)\langle\langle V\rangle\rangle$ was included into an 
$O(M_N^2/Q^2)$ kinematic Al power correction to the Bjp sum rule, which 
involves the second $x^2$ moments of the leading-twist contribution to 
$g_1^{p-n}=g_1^p-g_1^n$ and the  twist-3 matrix element, defined through the 
combination of $x^2$-weighted moments of the difference of 
structure functions  $ g_1^{p-n}$ and of 
$g_2^{p-n}=g_2^{p}(x,Q^2)-g_2^{n}(x,Q^2)$ 
as
\begin{equation}
\label{d2}
d_2^{p-n}=\int_0^1dx x^2\bigg(2g_1^{p-n}(x,Q^2)+3g_2^{p-n}(x,Q^2)\bigg)~~~,
\end{equation}
Taking into account this decomposition, it is possible to rewrite 
a theoretical expression for the numerator of the  $1/Q^2$ contribution, 
in the way  it was done say, in the most recent 
experimental  work of Ref.\cite{Deur:2004ti}   
\begin{equation}
\mu_4^{p-n}=\frac{M_N^2}{9}\big(a_2^{p-n}+4d_2^{p-n}+4f_2^{p-n}\big)~~,
\end{equation}
where 
\begin{equation}
a_2=\int_0^1dx x^2\big[g_1^{p}(x,Q^2)-g_1^{n}(x,Q^2)\big]
\end{equation}
is the target mass correction and 
\begin{equation}
2m_N^2f_2^{p-n}S_{\mu}=-4M_N S_{\mu}\langle\langle U^{\rm NS}
\rangle\rangle 
\end{equation}
is the twist-4 
contribution, which is related to the definition used by us as 
\begin{equation}
\langle\langle O_2 \rangle\rangle = \frac{1}{6}\frac{8}{9}\langle\langle 
U^{\rm NS}\rangle\rangle=-\frac{1}{6}\frac{4}{9}M_N^2f_2^{p-n} ~~.
\end{equation}
In other words we have the following relation 
\begin{equation}
M_N^2f_2^{p-n}=-2{\langle\langle U^{\rm NS}\rangle\rangle}~~.
\end{equation}
It should be stressed that in the region  where the perturbative theory 
is working well enough and the application of the 
operator-product expansion method is valid 
(say at $Q^2\geq 2~{\rm GeV}^2$),  both target mass corrections 
and twist-3 terms are small and we will neglect them in our further 
considerations \footnote{For completeness  
we note that there is a minor  difference  
between the the $O(M_N^2/Q^2)$ coefficients of the 
$\int_0^1 dxx^2 g_1^{p-n}$ terms in Ref.\cite{Balitsky:1989jb} and 
Ref.\cite{Deur:2004ti}. In the former and latter  cases
they are equal to (10/9) and 1 respectively.}.
These  features were revealed in the process of the analysis of Ref. 
\cite{Balitsky:1989jb}.

The third sum rule, which was originally derived for purely  theoretical 
purposes, is the Bjorken  unpolarized  sum rule\cite{Bjorken:1967px}. 
It can be written down as:
\begin{eqnarray}
\nonumber
{\rm Bjunp}(Q^2) &=& 
\int_0^1dx\bigg[F_1^{\nu p}(x,Q^2)-F_1^{\nu n}(x,Q^2)\bigg]
\\ \label{inp}
&=&C_{\rm Bjunp}(Q^2)- \frac{\langle\langle O_3 \rangle 
\rangle}{Q^2}-O\bigg(\frac{1}{Q^4}\bigg)~~~.
\end{eqnarray} 
It may  be also studied in future  as the valuable  test of
QCD both in perturbative and non-perturbative sectors.

As  in the  previous two cases, 
the coefficient function $C_{\rm Bjunp}(Q^2)$
is calculated up to next-to-next-to-leading order 
$a_s^3$-corrections\cite{Gorishnii:1983gs},\cite{Larin:1990zw}. 
The twist-4 matrix element 
to this sum rule was evaluated in Ref.\cite{Shuryak:1981kj}; with the 
following result:
\begin{equation}
\label{def}
\langle\langle O_3 \rangle \rangle =
\frac{8}{9}\langle\langle O^{\rm NS}\rangle\rangle~~~,  
\end{equation}
where the matrix element  $\langle\langle O^{\rm NS}\rangle\rangle$ 
is related to  the  
dimension-5 operator  
\begin{equation}
O_{\mu}^{\rm NS}=\overline{u}\tilde{G}_{\mu\nu}\gamma_{\nu}\gamma_5u-
\overline{d}\tilde{G}_{\mu\nu}\gamma_{\nu}\gamma_5d~~~,
\end{equation}
its matching over nucleon states 
\begin{equation}
\langle P|O_{\mu}^{\rm NS}|P\rangle=2p_{\mu}\langle\langle 
O^{\rm NS}\rangle\rangle~~~
\end{equation}
and application of Eq. (\ref{def}).

Let us now return to the  renormalon calculus.
The  basic theoretical problem  
is how to define the Borel image $B[D](\delta)$ (or the Borel sum)   
of the integral in  Eq. (\ref{Borel}) for the quantities we are 
interested in. In QCD 
this problem is usually solved using  perturbative methods and   
calculating the corresponding  multiloop Feynman diagrams with a   
one-gluon line, dressed by the chains of fermion 
bubbles (so called renormalon chain insertion). These chains are  
generating 
sign-alternating asymptotic perturbative series, typical of  the quantities 
under consideration, in powers   
of the expansion parameter  $N_f a_s$ (where $N_f$ is the number of quarks 
flavours).
  The contributions of these chains  are gauge-invariant, 
 but they  do not reflect
the whole picture of renormalon effects in QCD. The latter begin  
to manifest themselves after application of the naive non-abelianization
(NNA)  
ansatz\cite{Broadhurst:1994se}  only, namely after the 
replacement   $N_f\rightarrow -(3/2)\beta_0$ = $N_f-(33/2)$ in the leading 
terms of the large-$N_f$ expansion.
 This procedure  transforms a    
large-$N_f$ expansion  into  a  large-$\beta_0$ expansion, which 
in addition to quark  bubbles insertions into the  renormalon 
chain,  is taking 
into account the contributions of the  gluon- and ghost-bubbles insertions 
as well (though neglecting  definite   one-loop insertions into the 
gluon--quark--antiquark vertex, which should be also considered 
in the process of rigorous calculation  
of  the coefficient $\beta_0$). 
The application of the  NNA approach allowed the authors 
of Ref. \cite{Beneke:1994qe}
to formulate the extension to higher orders of the 
BLM-approach \cite{Brodsky:1982gc}. Technically, the work of Ref.
\cite{Beneke:1994qe}  supports the results of the 
first successful formulation of the BLM 
procedure to the  next-to-next-to-leading order \cite{Grunberg:1991ac}
Moreover, these two works pushed ahead the study of the 
BLM procedure in higher orders \cite{Mikhailov:2004iq}.
In principle, the relations of the results of Refs. 
\cite{Grunberg:1991ac, Beneke:1994qe,Mikhailov:2004iq}
need more detailed considerations. In view of the lack of space we will 
avoid discussions of this subjects here.

The Borel images calculated by this procedure  for the GLS and Bjp 
sum rules coincide and have the following form\cite{Broadhurst:1993ru}:
\begin{equation}
B[\rm C_{\rm Bjp}](\delta)=B[\rm C_{\rm GLS}](\delta)= 
-\frac{(3+\delta) {\rm exp}(5\delta/3)}{(1-\delta^2)(1-\delta^2/4)}~~.
\label{GLS}
\end{equation}
They contain  the IRR poles at $\delta=1$ and $\delta=2$ and the  
UVR poles at $\delta=-1$ and $\delta=-2$.
Note that the  $\delta=-1$ UVR poles in Eq. (\ref{GLS})
are suppressed by a factor $(1/2){\rm exp}(-10/3)=0.018$, relative 
to the dominant IRR poles at $\delta=1$ \cite{Broadhurst:2002bi}. Therefore,
in the  asymptotic structure of the perturbative QCD 
effects in the expressions for  
$C_{\rm GLS}(Q^2) \approx  C_{Bjp}(Q^2)$ 
(where  we neglect the small ``light-by-light-type'' effects, contributing to
$C_{\rm GLS}(Q^2)$) 
 the sign-constant part in Eq. (\ref{asymptotics}) dominates strongly 
with respect to the sign-alternating contribution, generated by 
$\delta=-1$ UVR. The scheme-dependence of these results are not so 
obvious, Indeed, the suppressions of $\delta=1$ UVR with respect to 
$\delta=1$ IRR is related to the application of the $\overline{\rm MS}$-scheme
which we are using  throughout the whole work.
In fact in this scheme the IRR renormalons  
are not suppressed.
However, 
there is the procedure, when the situation is reversed- 
the IRR are absent, but UVR may exist. This feature is 
manifesting itself 
for the models with ``frozen'' coupling constant (see e.g. 
\cite{Shirkov:1997wi} ).

Returning  to the large-$N_f$ expansion 
of the perturbative expressions 
\begin{equation}
 C_{\rm Bjp}(Q^2) =  C_{GLS}(Q^2)=1+\frac{C_F}{T_fN_f}\sum_{n=1}^
{\infty}r_n(T_fN_fa_s)^n~~,
\end{equation}
where $C_F=4/3$, $T_f=1/2$ and 
\begin{equation}
r_n={\rm lim}_{\delta\rightarrow 0}\bigg(-\frac{4}{3}\frac{ d}{d\delta}
\bigg)^{n-1}B[ C_{\rm Bjp}](\delta)~~,
\end{equation}
we arrive at  the following expansion in powers of $x=T_fN_fa_s$, namely   
\begin{equation}
\label{expansion}
\sum_{n}r_n x^n=-3x +8x^2-\frac{920}{27}x^3+\frac{38720}{243}x^4+...~~~,
\end{equation}
which is known in the $\overline{\rm MS}$ scheme up to  
$O(\alpha_s^9N_f^9)$ terms\cite{Broadhurst:1993ru}.
Using now   the traditional $\overline{\rm MS}$-scheme 
expansion in terms of the orders in $\alpha_s/\pi=4a_s$, one can  
compare the results of explicit perturbative  calculations of 
\begin{equation}
C_{\rm Bjp}(Q^2)=1+\sum_{n \geq 1}r_n\bigg(\frac{\alpha_s}{\pi}\bigg)^n
\end{equation}
with the known numbers 
\begin{eqnarray}
\label{r1}
r_1&=&-1 \\ \label{r2}
r_2&=&-4.5833+0.33333N_f \\ \label{r3}
r_3&=& -41.440+7.6073N_f-0.17747N_f^2
\end{eqnarray}
obtained at $O(\alpha_s^2)$ in Ref.\cite{Gorishnii:1985xm} 
and at $O(\alpha_s^3)$ in Ref.\cite{Larin:1991tj}, with the 
results of the  application of the    NNA 
procedure\cite{Broadhurst:1994se} to the estimates 
of the perturbative QCD corrections from large-$N_f$ expansion
of Eq. (\ref{expansion}) \footnote{It is worth noting  that  similar 
NNA analysis was  performed previously, in Ref.\cite{Lovett-Turner:1995ti}, 
for the $e^+e^-$ annihilation  Adler $D$-function.}. 
Performing the shift   $N_f\rightarrow 
N_f-33/2$ in the second, third and fourth  terms in Eq. (\ref{expansion}),  
we arrive at the following estimates in the $\overline{\rm{MS}}$ 
scheme \cite{Broadhurst:2002bi}:
\begin{eqnarray}
\label{r2NNA}
r_2^{\rm NNA}&=&-5.5 +0.33333N_f \\ \label{r3nna}
r_3^{\rm NNA}&=&-48.316+5.8565N_f-0.17747N_f^2\\ 
r_4^{\rm NNA}&=&-466.00+84.728N_f-5.1350~N_f^2+0.10374N_f^3~~~.
\end{eqnarray}
 Reasonable  agreement can be observed  between the sign structure 
and values of the NNA estimates  
and the results of explicit calculations 
(compare the estimates of Eqs. (\ref{r2NNA}) and (\ref{r3nna})
with the numbers in Eqs. (\ref{r2}) and  (\ref{r3}), respectively).  
As to the prediction for $r_4^{\rm NNA}$, it may serve as a  guide 
for understanding the rate of growth of the coefficients of the 
perturbative series generated by the  single 
renormalon-chain approximation.

Consider now  the Bjunp sum rule, which is  defined in Eq. (\ref{inp}). 
Within the large-$N_f$, expansion its perturbative 
coefficient function 
\begin{equation}
C_{\rm Bjpunp}(Q^2)=1+\sum_{n\geq 1}\tilde{r}_n
\bigg(\frac{\alpha_s}{\pi}\bigg)^n~~~
\end{equation}
was calculated in the $\overline{\rm MS}$ scheme and  large-$N_f$ expansion  
up to a  
$O(\alpha_s^9N_f^9)$-terms\cite{Broadhurst:2002bi}.
Following the logic of our work, we present here 
the results for the   first 4 terms only:
\begin{equation}
\label{expansion2}
\sum_{n}\tilde{r}_n x^n=-2x +\frac{64}{9}8x^2-\frac{2480}{81}x^3+
\frac{113920}{729}x^4+ \dots 
\end{equation}
As was already mentioned above, the explicit results 
of calculations of the perturbative contributions to the 
Bjunp sum rule 
\begin{equation}
C_{\rm Bjunp}(Q^2)=1+\sum_{n\geq 1}\tilde{r}_n
\bigg(\frac{\alpha_s}{\pi}\bigg)^n
\end{equation}
are known up to the  order $O(\alpha_s^3)$ level. These 
results are: 
\begin{eqnarray}
\label{tilder1}
\tilde{r}_1&=&-2/3 \\
\label{tilder2}
\tilde{r}_2&=& -3.8333+0.29630N_f \\ \label{tilder3}
\tilde{r}_3&=& -36.155+6.3313N_f-0.15947N_f^2
\end{eqnarray}
where $\tilde{r}_2$ was calculated in Ref.\cite{Gorishnii:1983gs}
while  $\tilde{r}_3$ was evaluated in Ref.\cite{Larin:1990zw}.
Applying now the NNA procedure to the  
series of Eq. (\ref{expansion2}), 
we find that, in the $\overline{\rm{MS}}$ scheme,
the estimated  coefficients of the
Bjunp sum rules have the following form\cite{Broadhurst:2002bi}:  
\begin{eqnarray}
\label{2NNA}
\tilde{r}_2^{\rm NNA}&=&-4.8889 +0.29630N_f \\ \label{3nna}
\tilde{r}_3^{\rm NNA}&=&-43.414+5.2623N_f-0.15947N_f^2\\ 
\tilde{r}_4^{\rm NNA}&=&-457.02+83.094N_f-5.0360~N_f^2+0.10174N_f^3~~~.
\end{eqnarray}
The estimate of Eq. (\ref{2NNA})
is in agreement with its exact partner of Eq. (\ref{tilder2}).
The same  situation holds for the  $O(\alpha_s^3)$ 
corrections (compare Eq. (\ref{3nna}) with Eq. (\ref{tilder3})).
It should be stressed, that the similarity of the next-to-next-to-
leading-order 
$\overline{\rm {MS}}$-scheme  
perturbative QCD contributions to the Bjp and Bjunp sum rules  
was previously noticed  
in Ref.\cite{Gardi:1998rf}, although no explanation of this 
observation was given. Now,  within the NNA procedure, 
 it is possible  to generalize this 
observation  to higher-order level. Indeed, the NNA estimates 
of the $O(\alpha_s^4)$ corrections to the Bjp and Bjunp sum rules 
have a  similar expressions as well. 
{\bf  These facts may indicate  the close 
similarity in the full perturbative structure of the QCD corrections 
to the Bjunp sum rule, the Bjp sum rule and the GLS sum rule}  
(provided the  ``light-by-light-type''
terms will  not drastically modify the values of perturbative 
terms  in  the latter case in the one-renormalon chain approximation).  
Note  that, generally speaking,  from this order of perturbation theory 
the diagrams from the 
second renormalon chain are starting to contribute to the quantities 
under consideration. 
These diagrams may influence the asymptotic behavior of the  
the series  considered\cite{Vainshtein:1994ff}. In view of this 
it seems that it is  more rigorous to use, in the 
phenomenological 
application, the order of  $\alpha_s^4$-terms, estimated 
in Ref.\cite{Kataev:1995vh} using the  PMS approach\cite{Stevenson:1981vj}
and  the effective-charges approach, developed in Ref.\cite{Grunberg:1982fw}.
However, since in this work we concentrated ourselves on the 
structure of the QCD expressions, obtained in the one-renormalon chain 
approximation, we will avoid more detailed discussions of the possible 
influence of the multi-renormalon chain contributions to 
the results of our studies.

The  observed  in Ref.\cite{Gardi:1998rf} 
similarity of the  next-to-next-to-leading-order approximations 
for the Bjp and Bjunp sum rules 
was attributed in Ref.\cite{Broadhurst:2002bi}
to the fact that  the dominant $\delta=1$  IRR   contribution  to the 
Borel images of these sum rules enters with   identical residues.
Indeed, 
the Borel images in the Borel integrals of Eq. (7) for the Bjunp and 
Bjp sum rules  turn out to be  closely  
related\cite{Broadhurst:2002bi}, namely   
\begin{equation}
\label{rens}
B[ C_{\rm Bjunp}](\delta)=\bigg(\frac{2(1+\delta)}{3+\delta}\bigg)
B[C_{\rm Bjp}](\delta)=- \frac{2{\rm exp}(5\delta/3)}
{(1-\delta)(1-\delta^2/4)})~~~.
\end{equation}  
Comparing Eq. (27) with Eq. (\ref{rens}) one can convince oneself 
that the residues of the poles at $\delta=1$ in these two 
expressions are really the same and are equal to the factor 
$-(8/3){\rm exp}(5/3)$.

Notice also  the {\bf absence} of $\delta=-1$  UVR pole and 
the {\bf existence} in Eq. (\ref{rens}) of a $\delta=-2$ UVR pole  
together with  the 
leading  $\delta=1$ IRR one. Thus we are observing 
one more  interesting fact:   
the structure of the Borel image for the Borel sum, related to the 
 Bjunp sum rule,  
 is {\bf dual} to the structure of leading renormalon contributions 
to the Borel image of the Borel sum for the 
 $e^+e^-$ annihilation Adler D-function. 
Indeed, in the latter case the leading IRR is manifesting itself at 
$\delta=2$, while the leading UVR pole is appearing at $\delta=-1$
(the general structure of  renormalon
singularities in the   $e^+e^-$ annihilation channel was analyzed  
in  Ref.\cite{Zakharov:1992bx}, while the 
concrete $\overline{\rm MS}$-scheme  calculations of the corresponding Borel 
image  were done later on  in Refs.\cite{Beneke:1992ch} and 
\cite{Broadhurst:1992si}). 

The absence of $\delta=1$ IRR  
in the Borel sum of the  $e^+e^-$ annihilation channel is related 
to the absence of  $O(\Lambda^2/Q^2)$ non-perturbative power correction 
in the standard variant of the operator product expansion formalism, applied to
the theoretical expression for the $e^+e^-$ annihilation 
 Adler D-function. 
Indeed, the existence of  lowest dimension-4 quark and gluon 
condensates \cite{Shifman:1978bx} 
in this channel 
can be associated in terms of  renormalon language with the existence 
of the {\bf leading} IRR pole, which in case of 
``Borelization'' of the Adler D-function 
is appearing at $\delta=2$. However, as
was already discussed above,
the dimension-2 non-perturbative corrections enter into the theoretical 
expressions for the three DIS sum rules we are interested in. In  the IRR 
language, this corresponds to the appearance of a $\delta=1$ 
IRR pole\cite{Mueller:1993pa}, which manifests itself in the  concrete 
results of Refs.\cite{Broadhurst:1993ru},\cite{Broadhurst:2002bi} 
(see Eqs. (27) and  (47)). Thus, it should be stressed that 
the structure of singularities of the Borel sums (or images) 
is not universal and depends from the physical quantity under 
consideration. 

\section{IRR for DIS sum rules  and the values of twist-4 corrections}
In addition to controlling the sign-positive $n!$ growth  of the 
asymptotic series the existence of $\delta=1$ IRR 
gives an ambiguity in taking the  Borel integral of Eq. (7)
over this pole. In the case of large $\beta_0$ expansion 
and for the series we are interested in, this ambiguity 
was estimated in Ref.\cite{Beneke:2000kc}. 
Moreover, $\delta=1$ IRR 
generates 
the negative power suppressed correction which  
has the following expression: 
\begin{equation}
\label{ambig}
\Delta  C_{\rm sum~rules}\approx -\frac{32 \rm exp(5/3)}
{9 \beta_0}\frac{\Lambda^2_{\overline{\rm MS}}}{Q^2}~~.
\end{equation}
Notice, that  it has the same negative sign as the residue of 
$\delta=1$ IRR.

This estimate   
may  be coordinated 
with the  definition of the twist-4 
matrix element in the sum rules we are interested in.
Therefore,  we will make  the   assumption that 
the identical values and signs of the IRR induced power-suppressed 
term  indicate  
that the values of twist-4 contributions to the 
expressions of GLS, Bip and Bjunp sum rules, normalized to unity,  
should have the same negative sign and 
a similar  closed value  \cite{Kataev:2005ci}. 
This assumption is similar to the known in the renormalon-oriented 
literature guess of ``universality''\cite{Dokshitzer:1995qm}.

Let us check this assumption, considering the following 
expressions for the sum rules we are interested in 
\begin{eqnarray}
{\rm GLS}(Q^2)& =&3\bigg[1-4a_s-O(a_s^2) -
\label{A}
\frac{\rm A}{Q^2}\bigg]~~~, \\ \label{Bp}
{\rm Bjp}(Q^2)&=&\frac{g_A}{6}\bigg[1-4a_s-O(a_s^2)-\frac{\rm {B} }{Q^2}
\bigg]~~, \\ \label{C}
{\rm Bjunp}(Q^2)&=&\bigg[1-\frac{8}{3}a_s -
O(a_s^2)-\frac{ \rm C}{Q^2}\bigg]~~~,
\end{eqnarray}
where ${\rm A} = \langle\langle O_1 \rangle\rangle/3$, 
${\rm B}= \langle\langle O_2\rangle\rangle(6/g_A)$ and ${\rm C}= 
\langle\langle O_3\rangle\rangle$
 and compare in Table 1 the results of different 
theoretical and phenomenologically based evaluations 
of the twist-4 parameters ${\rm  A},~{\rm B}$ and ${\rm C}$. 

\begin{table}[t]
  \centering
  \caption{ \it The results for twist-4 contributions  
to  the GLS, Bjp and Bjunp sum-rule expressions of Eqs. (49)--(51).
    }
  \vskip 0.1 in
  \begin{tabular}{|l|c|c|c|} \hline
    &  ${\rm A}~[{\rm GeV}^2]$ & ${\rm B}~[{\rm GeV}^2]$ & 
${\rm C}~[{\rm GeV}^2]$ \\
    \hline
    \hline
    QCD sum rules (Ref.\cite{Braun:1986ty}) & $ 0.098  
\pm 0.049$     &  ---
    & $ 0.133 \pm 0.065$ \\
    QCD sum rules (Ref.\cite{Balitsky:1989jb} & --- 
& $0.063
 \pm 0.031$ & --- \\
    QCD sum rules (Ref.\cite{Ross:1993gb} &    
$0.158 \pm 0.078~$ &    $0.223  \pm 0.118~$  & 
                $0.16 \pm 0.08~$            \\
QCD sum rules (Ref.\cite{Stein:1994zk}) & ---   &  
$0.025 \pm 0.012~$       & ---
 \\
    Instanton model (Ref.\cite{Balla:1997hf}) 
& $0.078 \pm 0.039~$ & $ 0.087 \pm 0.043$ & ---
\\
  Instanton model (Ref.\cite{Weiss:2002vv}) & --- & --- 
& $0.16 \pm 0.08 $ \\    
Experiment (Ref.\cite{Sidorov:2004sg}) & ----& $0.098 \pm 0.028$ &---\\  
Experiment (Ref.\cite{Sid}) & $0.04 \pm 0.13$ & ---& ---\\
\hline
\end{tabular}
\end{table}
In the case of the GLS and Bjunp sum rules  the results of  
the original application of the three-point function QCD sum rules method 
gave $\langle\langle O^{\rm S}\rangle\rangle = 0.33~{\rm GeV^2}$ 
and $\langle\langle O^{\rm NS}\rangle\rangle = 0.15~{\rm GeV^2}$, 
with over 50$\%$ error bars\cite{Braun:1986ty},
while  the three-point function estimates 
for the modified results of calculations 
of the twist-4 parameter of the  Bjp sum rule 
resulted in the following value  
$M_N^2f_2^{p-n} = - 0.18 \pm 0.09~ {\rm GeV^2}$
in the region where nucleon target mass corrections of  $O(M_N^2/Q^2)$ 
and twist-3 contribution may be neglected\cite{Balitsky:1989jb}. 
As was already mentioned, these calculations 
were re-examined  using three-point function QCD sum-rules approach 
in Refs.\cite{Ross:1993gb} and \cite{Stein:1994zk}.
In the first case the obtained result turned out to 
be larger than the original results from Ref.\cite{Balitsky:1989jb} 
and has the following value of  
$M_N^2f_2^{p-n} = - 0.634 \pm 0.317~ {\rm GeV^2}$
\cite{Ross:1993gb}, while in the  latter case it was considerably smaller,
namely $M_N^2f_2^{p-n}=-0.07\pm 0.035$ \cite{Stein:1994zk}, 
although within $50\%$ theoretical uncertainty  we adopt 
for all calculations within three-point function QCD  sum-rules approach, 
this value does not disagree with the results obtained in 
Ref.\cite{Balitsky:1989jb}. The relatively high difference between 
the central values of 
estimates  of Refs.\cite{Ross:1993gb} and \cite{Balitsky:1989jb}
is explained by the fact that in the former analysis  the additional 
corrections to the perturbative side of the corresponding QCD sum rules are 
included and the continuum term to the nucleon pole of the low-energy 
side of this sum rule is explicitly retained. This leads to 
better stability of the extracted value of matrix element with 
respect to the Borel parameter and increases its central value. 
Note, however, that the theoretical error of the three-point 
function of the  QCD sum rules result 
of Ref. \cite{Ross:1993gb} is considerably underestimated.
We fix it as 50 $\%$ uncertainty, which to our point 
of view is  typical to 
all three-point function QCD sum rules results.

In Table 1 we present the estimates of twist-4 
corrections to different DIS sum rules, obtained 
with the help of the  three-point function QCD sum-rules approach and 
compare  them with the results of  the application    
of different  theoretical approach, based 
on the picture of the QCD vacuum as a ``medium'' of 
instantons\cite{Shuryak:1981ff}. This picture was  further developed in 
the method in Ref.\cite{Diakonov:1983hh} and applied for  estimating   
twist-4 contributions to the GLS sum rule and  Bjp sum rule 
 in Ref.\cite{Balla:1997hf}, while the number for the twist-4 
contribution to the Bjunp sum rule, which follows from this approach, 
 was presented in Ref.\cite{Weiss:2002vv}. 
In the absence of estimates 
of theoretical uncertainties within this approach, we will 
apply to them the careful $50\%$ estimate as well.  All these results 
support the original results of the  three-point function QCD sum 
rules calculations 
of the twist-4 corrections to the GLS, Bjunp sum rules\cite{Braun:1986ty}
and Bjp sum rule\cite{Balitsky:1989jb}, though the additional 
three-point function QCD sum rules cross-check of the results 
of Ref.\cite{Balitsky:1989jb} may be rather useful.

The  experimentally 
motivated value of the twist-4 contribution  to the Bjorken sum rule 
$M_n^2f_2^{p-n}=-0.28 \pm 0.08~{\rm GeV}^2$ \cite{Sidorov:2004sg}
was obtained  by means of  integrating in $x$   
the numerator of the dimensionless  
$ h(x)/Q^2$ contributions, extracted from the fits of  
world average data for $g_1^{p}(x,Q^2)$ and $g_1^{n}(x,Q^2)$ performed in 
Ref. \cite{Leader:2002ni}. From the results of Table 1 one can see 
that the agreement with the QCD sum-rules calculations 
of Ref.\cite{Balitsky:1989jb} and instanton-based calculations 
of Ref.\cite{Balla:1997hf} is more than qualitative.
 
The experimentally inspired  estimate  for the 
twist-4 contribution to the GLS sum rule was obtained only recently\cite{Sid}
as a result of the integration of $x$-dependence 
of the twist-4 contribution $h(x)/Q^2$, extracted in the works of 
 Ref.\cite{Kataev:1997nc} devoted to the  
analysis of $xF_3$ data of CCFR collaboration. One can see that the 
central value of the contribution  is negative 
(in fact it comes with the negative sign in the sum rule) 
but has 
rather large uncertainties. So, at the present level we cannot 
obtain from this estimate even qualitative information and 
additional work on its improvement is needed.   

To conclude, we present the final results for the 
GLS, Bjp and Bjunp sum rules, where for definiteness 
the twist-4 matrix elements are estimated using 
the central values of the three-point function QCD sum-rules 
results from  
Refs.\cite{Braun:1986ty} and \cite{Balitsky:1989jb}: 
\begin{eqnarray}
{\rm GLS}(Q^2)& =&3\bigg[1-4a_s-O(a_s^2) -
\frac{0.098~{\rm GeV^2}}{Q^2}\bigg]~~~, \\
{\rm Bjp}(Q^2)&=
&\frac{g_A}{6}\bigg[1-4a_s-O(a_s^2)-\frac{0.063~{\rm GeV^2}}{Q^2}
\bigg]~~, \\
{\rm Bjunp}(Q^2)&=&\bigg[1-\frac{8}{3}a_s -
O(a_s^2)-\frac{0.133~{\rm GeV^2}}{Q^2}\bigg]~~~.
\end{eqnarray}

It should be stressed  that they all have the same negative sign 
and within existing  theoretical 
uncertainties are in agreement with each other. This fact was 
anticipated  by the identical value of the ambiguity, generated 
by the $\delta=1$  IRR pole of the Borel images  of all 
these three sum rules (see Eq. (48)). Moreover, 
as  follows from the  
results of application of the single-renormalon 
chain approximation in the perturbative sector presented in Sec.2, 
we may  expect a similar 
asymptotic behavior of the perturbative corrections to all 
these three sum rules (compare Eqs.(33)--(36) with Eqs.(42)--(45)).
It is interesting  that the similar property is manifesting itself 
in perturbative series under investigations 
at the $O(\alpha_s^3)$ level, studied within scheme-invariant approaches 
in Ref.\cite{Kataev:1995vh}. 

These  facts  give us the idea that the sum rules we are interested in 
are closely related and that, 
in the region, where we can neglect target mass corrections and twist-3 
contributions to the Bjp  sum rule and quark-mass dependent corrections 
(say in the region $Q^2 \geq 2~{\rm GeV}^2$) 
we can write down the following basic 
relation \cite{Kataev:2005ci}:
\begin{equation}
{\rm Bjp}(Q^2)\approx (g_A/18){\rm GLS}(Q^2)\approx (g_A/6){\rm 
Bjunp}(Q^2)~~~.
\label{basic}
\end{equation}

In the next section we will present more detailed considerations 
of the experimental consequences of these relations then 
those, that are briefly outlined in Ref.\cite{Kataev:2005ci}.

\section{IRR- inspired relations and experiment}
In order to test whether our basic relation Eq. (\ref{basic}) is respected 
by experiment, we first present the results of  the extraction of the 
GLS sum rule by combining CCFR neutrino DIS data with the data 
for other neutrino DIS experiments for $1~{\rm GeV}^2 < Q^2 
< 15 ~{\rm GeV^2}$ \cite{Kim:1998ki}. 
It is known that the  weighted  extraction of 
$\alpha_s(M_Z)$ from these  data result in the  rather rough value 
$\alpha_s(M_Z)= 0.115\pm^{0.009}_{0.12}$, which,          
 is in agreement 
with $\alpha_s(M_Z)=0.115 \pm 0.001~(stat) \pm 0.005~(syst)$ $\pm 
0.003~(twist)\pm 0.0005~(scheme)$, extracted in Ref.\cite{Chyla:1992cg}
from the previous CCFR data for the GLS sum rule at 
$Q^2=3~{\rm GeV}^2$\cite{Leung:1992yx}. 
However, for our purposes we will not need to re-extract $\alpha_s(M_Z)$ 
values from the GLS sum rule results of Ref.\cite{Kim:1998ki},
but will use these, 
which are presented in Table 2.  
\begin{table}[t]
  \centering
  \caption{ \it The results for the GLS sum rule from Ref. \cite{Kim:1998ki}
}
  \vskip 0.1 in
  \begin{tabular}{|l|c|c|} \hline
   $Q^2$ [${\rm GeV^2}$]  &   GLS sum rule \\
    \hline
    \hline
    2.00   & $2.49 \pm 0.08 \pm 0.14$ \\ 
    3.16  &  $2.55 \pm 0.08  \pm 0.10$ \\
    5.01  &  $2.78 \pm 0.06 \pm 0.19$                     \\
    7.94  &  $2.82 \pm 0.07 \pm 0.19$ \\
    12.59 &  $2.80 \pm 0.13 \pm 0.18$ \\ 
\hline
  \end{tabular}
\end{table}

To estimate the values of the Bjp sum rule from the results 
of Table 2 we will use our main equation (\ref{basic}) 
 and will compare them with available experimental 
data for the Bjp sum rule. The results of these studies are presented in 
Table 3.

\begin{table}[t]
  \centering
  \caption{ \it The comparison of the results of application
of Eq. (\ref{basic}) with direct experimentally motivated  numbers}
  \vskip 0.1 in
  \begin{tabular}{|l|c|c|} \hline
$Q^2$ [${\rm GeV^2}$]   &  Bjp from  Table 1   & Bjp SR   (exp) \\
    \hline
    \hline
    2.00   & $0.174 \pm 0.006 \pm 0.010$ & $0.169\pm 0.025$~ 
[Ref.\cite{Abe:1998wq}] \\ 
    3.16  &  $0.178 \pm 0.004  \pm 0.007$ & $0.164 \pm 0.023$ 
~[Ref.\cite{Abe:1998wq}] \\
    5.01  &  $0.195 \pm 0.004\pm 0.013$ & $0.181   \pm 
0012~(stat)\pm 0.018~ (syst)$~[Ref.\cite{Adeva:1998vw}] \\
    7.94  &  $0.197 \pm 0.005 \pm 0.013$ & -----  \\
    12.5   & $0.196 \pm 0.009  \pm 0.013$ & $0.195 \pm 0.029$  
Ref.~\cite{Adeva:1997is}  \\
    \hline
  \end{tabular}
\end{table}
 
One can see that though the central values 
of estimated numbers for the Bjp SR are higher
than the results of the SLAC E143 collaboration \cite{Abe:1998wq}, they 
agree  within error bars. It is also  interesting to compare  
the result from Table 3 with the value of the Bjp sum rule 
extracted in Ref.\cite{Altarelli:1996nm} from the SLAC and SMC data 
${\rm Bjp}(3~{\rm GeV^2})=0.177 \pm 0.018$ and which, within error bars, 
do not contradict the value ${\rm Bjp}(3~{\rm GeV^2})=0.164 \pm 0.011$
used in the work of Ref.\cite{Ellis:1995jv}. It is rather inspiring that 
within error bars these results agree with the GLS sum rule value 
at $Q^2= 3.16~{\rm GeV^2}$. The same feature holds for  the Bjp sum rules 
at $Q^2= 5~{\rm GeV^2}$, namely for the SMC result of Ref.\cite{Adeva:1998vw}.
Thus we think 
that within existing  uncertainties our approximate IRR-inspired  
basic equation (\ref{basic}) is supported by existing experimental data.

\section{Conclusions}
We demonstrated  that the existing phenomenological data 
do not contradict  the   basic relation of Eq. (\ref{basic}) and therefore  
{\bf the reliability of the  one-renormalon chain approximation of 
the theoretical 
quantities under consideration}. For its more  detailed studies, we 
may rely on the appearance of Neutrino Factory data for all sum rules,
which enter in Eq. (\ref{basic}). In fact it may provide rather useful 
data not only for the GLS and Bjp sum rules, but for the Bjunp 
sum rule as well (for a discussion of this possibility  see 
Refs.\cite{Mangano:2001mj},\cite{Alekhin:2002pj}).  

Another interesting option of the relation of Eq. (\ref{basic}) is 
to analyze the sources of its possible violation 
in the lower
energy region of over $Q^2\approx 1~{\rm GeV^2}$, where one 
may compare the CCFR data for the GLS sum rule at  
the energy point $Q^2=1.26~{\rm GeV^2}$ 
 \cite{Kim:1998ki} and the JLAB data for the Bjp sum rule 
at $Q^2=1.10~{\rm GeV}^2$ \cite{Deur:2004ti}.

To conclude this section, we would like to emphasize that the 
problems considered by us in this work  are complementary to
the considerations of Ref.\cite{Brodsky:1995tb}. In the former 
analysis,  
the GLS and Bjp sum rules were determined in the high
energy point of over $Q^2=12.33~{\rm GeV}^2$
from the   
generalized Crewther relation constructed in \cite{Broadhurst:1993ru},
\cite{Brodsky:1995tb},\cite{Crewther:1972kn}
using the  extension of the  
BLM approach of Ref.\cite{Brodsky:1982gc} and the analysis of $e^+e^-$ 
annihilation data from Ref.\cite{Mattingly:1993ej}.
Certainly, the renormalon-chain insertions are absorbed in this 
approach into the BLM scale. However, the considerations within this 
language of the high-twist effects 
is still missed. It may be of interest to think of the possibility 
of evaluating high-twist contributions to the Crewther relation, 
which relates, in the Eucledian region we are working  
massless QCD perturbative contributions to the  Adler D-function 
of $e^+e^-$-annihilation  with the perturbative 
corrections to the   GLS and Bjp sum rules.

{\bf Acknowledgements}
I am grateful to D.J. Broadhurst for a productive collaboration.
It is a pleasure to express my personal thanks to S.I. Alekhin, G. Altarelli,
Yu.L. Dokshitzer,
J. Ellis, G. Grunberg, A.V. Sidorov and V.I. Zakharov for useful 
discussions at  various stages of this work. 
 This article  grew up from my talk at the 
 19th Rencontre de Physique de la Vall\`ee d'Aoste
(27 February-5 March,2005,  La Thuile, Aosta Valley, Italy).
I would like to thank its organizers, G. Bellettini and M. Greco 
for their  invitation.
This work is supported by RFBR Grants N03-02-17047, 03-02-17177 and  
N 05-01-00992.
It was continued  during the visit  to CERN. I have  real pleasure 
in thanking the members of the CERN 
Theory Group for hospitality. In its final form the work was 
completed during the visit to ICTP (Trieste). I am grateful 
to a referee for constructive advise
\end{document}